\documentclass[prb,aps,twocolumn,draft,tightenlines,%
showpacs,floatfix,superscriptaddress]{revtex4}
\usepackage{psfig}
\usepackage{amssymb}
\usepackage{psfig}
\usepackage{amsmath}
\usepackage{colordvi}
\begin{document}

\title{Spin-dependent hole quantum transport in Aharonov-Bohm
ring structure: possible schemes for spin filter}

\author{J. Zhou}
\affiliation{Hefei National Laboratory
for Physical Sciences at
  Microscale, University of Science and Technology of China,
Hefei, Anhui, 230026, China}
\affiliation{Department of Physics, University of Science and
Technology of China, Hefei, Anhui, 230026, China}
\altaffiliation{Mailing Address.}
\author{M. W. Wu}
\thanks{Author to whom all correspondence should be addressed}%
\email{mwwu@ustc.edu.cn} 
\affiliation{Hefei National Laboratory
for Physical Sciences at
  Microscale, University of Science and Technology of China,
Hefei, Anhui, 230026, China}
\affiliation{Department of Physics, University of Science and
Technology of China, Hefei, Anhui, 230026, China}
\altaffiliation{Mailing Address.}
\author{M. Q. Weng}
\affiliation{Department of Physics and Engineering Physics, Stevens 
Institute of Technology, Hoboken, NJ 07030, USA}
\date{\today}

\begin{abstract}
We study the Aharonov-Bohm (AB) effect in two-dimensional mesoscopic 
frame in hole systems. We show that differing from the AB
effect in electron systems, due to the presence of both the heavy 
hole and the light hole, the conductances not only show 
the normal spin-unresolved
AB oscillations, but also become spin-separated. 
Some schemes for spin filter based on the abundant 
interference characteristics are proposed and the robustness against
the disorder of the proposed schemes is discussed.
\end{abstract}

\maketitle

The aim of using not only charge but also spin degree of freedom
of electrons and holes in semiconductor electronic devices leads
to a new field: semiconductor spintronics.\cite{prinz} Spin
filter is one of the basic devices in this field. 
Many schemes for spin filters, most in
electron systems, have been proposed\cite{filter} in order to
inject spin-polarized current into semiconductors, by means of
spin-selective barriers, stubs,\cite{stub} weak periodic
magnetic modulations\cite{wu,wu1} and anti-resonance effects in a double-bend
structure.\cite{wu2}

In this paper, we study the AB effect\cite{ab,wu1} 
in two-dimensional mesoscopic hole system.
The interferences between the four spin states, {\em i.e.},
the spin-up and -down heavy hole (HH) states and the spin-up and -down
light hole (LH) states are more complicated than the electron 
system. Possible schemes for
spin filter  are proposed based on the abundant interference
characteristics: When the Fermi energy of the lead is
lower than the LH band edge of the frame,
one can use the AB frame as a spin filter of HH by controlling the
AB flux. When a suitable strain is applied on the frame  to 
make the band edges of the HH and the LH close to each other,
then if one injects a spin unpolarized HH
current into the frame, a spin polarized LH (or HH) current  can be obtained by
controlling the AB flux. 

\begin{figure}[htb]
\vskip-0.3cm
\centerline{\psfig{figure=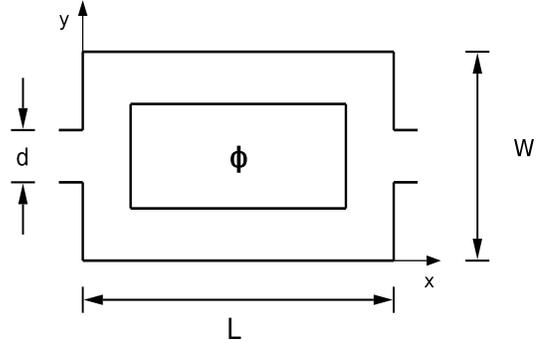,width=7cm,height=5cm,angle=0}}
\vskip -0.3cm
  \caption{Schematic view of a AB frame structure in 2D hole system with arm width $d$,
  frame length $L$ and frame width $W$.
}
\label{fig:2dring}
\end{figure}

We consider the AB flux $\phi$ introduced by a homogeneous
magnetic field $B$ through a two-dimensional (2D) AB frame
structure as shown in Fig.\ \ref{fig:2dring}, which is grown
in a (001) GaAs quantum well with a small well width 
($a=\sqrt{10}$\ nm). The
momentum states along the growth direction ($z$) are therefore
quantized and one only need to consider the lowest subband. In
this system there is no spin correlation $\langle
a_{k\frac{3}{2}}^{\dag}a_{k-\frac{3}{2}}\rangle$ ($\langle
a_{k\frac{1}{2}}^{\dag}a_{k-\frac{1}{2}}\rangle$) between the
spin-up and -down HH's (LH's). The spin-up HH's (LH's) are only
coupled with the spin-down LH's (HH's). This can be seen from the
Luttinger Hamiltonian\cite{trebin} $H_L$ in the momentum space
with the matrix elements arranged in the order of $\frac{3}{2}$,
$\frac{1}{2}$, $-\frac{1}{2}$ and $-\frac{3}{2}$:
\begin{equation} H_{L} =\frac {1}
{2m_0}\left(\begin{array}{cccc}
P+Q & 0 & R & 0 \\
0 & P-Q & 0 & R \\
R^\dag & 0 & P-Q & 0 \\
0 & R^\dag & 0 & P+Q
\end{array}\right)\ .
\end{equation}
In this equation $P\pm Q=(\gamma_1\pm\gamma_2)(P_x^2+P_y^2)+
(\gamma_1\mp 2\gamma_2)\frac{\pi^2}{a^2}|t|$ and
$R=-\sqrt{3}[\gamma_2(P_x^2-P_y^2)-2i\gamma_{3}P_{x}P_{y}]$ with
$\gamma_1=6.85$ and $\gamma_2=2.1$
 representing the Luttinger coefficients.\cite{strain}
$m_0/(\gamma_1\pm \gamma_2)$ are the effective
masses of the HH and the LH in the $x-y$ plane with $m_0$ representing the free electron mass.
Additionally, the Luttinger Hamiltonian can be separated into two independent parts:
$ H_{\alpha}(k_x,k_y) =\frac {1}
{2m_0}\left(\begin{array}{cccc}
P+Q & R \\
R^{\dag} & P-Q \\
\end{array}\right)$, with the matrix elements arranged in the order
of $\frac{3}{2}$ and $-\frac{1}{2}$
for the spin-up HH and the spin-down LH subsystem noted as $\alpha$,
and $ H_{\beta}(k_x,k_y) =\frac {1}
{2m_0}\left(\begin{array}{cccc}
P+Q & R^{\dag} \\
R & P-Q \\
\end{array}\right)$, with the matrix elements arranged in the order
of $-\frac{3}{2}$ and $\frac{1}{2}$
for the spin-down HH and the spin-up LH subsystem noted as $\beta$.

In real space, the Hamiltonian with AB flux can be written in the tight-binding version as:
\begin{widetext}
\begin{eqnarray}
  &&H_{2D}=\sum_{i,j,\sigma=\pm\frac{3}{2},\pm\frac{1}{2}}
  \epsilon_{\sigma}a_{i,j,\sigma}^{\dag}a_{i,j,\sigma}
  +\sum_{i,j,\sigma=\pm\frac{3}{2},\pm\frac{1}{2},\delta=\pm1}
  (\gamma_1\pm\gamma_2)V_{i^\prime j^\prime,ij}
  [a_{i+\delta,j,\sigma}^{\dag}
  a_{i,j,\sigma}+a_{i,j+\delta,\sigma}^{\dag}a_{i,j,\sigma}] \nonumber \\
  &&+\Big\{\sum_{i,j,\delta=\pm1,\lambda=0,1}(-\sqrt{3})\gamma_{2}V_{i^\prime j^\prime,ij}[a_{i+\delta,j,
\frac{3}{2}-\lambda}^{\dag} a_{i,j,-\frac{1}{2}-\lambda}-
  a_{i,j+\delta,\frac{3}{2}-\lambda}^{\dag}a_{i,j,-\frac{1}{2}-\lambda}] \nonumber \\
  &&+\sum_{i,j,\delta=\pm1,\lambda=0,1}\frac{\sqrt{3}}{2}i\gamma_{3}V_{i^\prime j^\prime,ij}
  [a_{i+\delta,j+\delta,
\frac{3}{2}-\lambda}^{\dag} a_{i,j,-\frac{1}{2}-\lambda}-
  a_{i+\delta,j-\delta,\frac{3}{2}-\lambda}^{\dag}a_{i,j,-\frac{1}{2}-\lambda}]
  +\mbox{H.C.}\Big\}
\end{eqnarray}
\end{widetext}
where $i$ and $j$ denote the coordinates along the $x$- and $y$-axes.
$t=-\hbar^2/{(2m_{0}a_0^2)}$ is the energy unit with $a_0$ standing for the
``lattice'' constant.
With the vector potential ${\bf A}$ in the Laudau gauge, {\em i.e.},
${\bf A}=(-\frac{1}{2}By,\frac{1}{2}Bx,0)$, the
hopping energy from ${\bf r}_{i,j}$ to ${\bf r}_{i^\prime,j^\prime}$ 
is given by
$V_{i^\prime j^\prime,ij}=t\exp[ie{\bf A}\cdot ({\bf r}_{i^\prime, j^\prime}
-{\bf r}_{i,j})/\hbar]$.
 $\epsilon_{\pm\frac{3}{2}}
=(\gamma_1- 2\gamma_2)\frac{\pi^2}{a^2}|t|-(\gamma_1+ \gamma_2)4t$
and $\epsilon_{\pm\frac{1}{2}}
=(\gamma_1+ 2\gamma_2)\frac{\pi^2}{a^2}|t|-(\gamma_1- \gamma_2)4t$
with the first terms standing for the lowest subband
energy in the $z$ direction. The first and the second terms in $\{\cdots\}$
are the nearest-neighbor and the next-nearest-neighbor spin-flip hopping terms.
Obviously, there is not any direct or indirect spin flip between the spin-up and
-down HH's or between the spin-up and -down LH's in the Hamiltonian.
Additionally
\begin{equation}
H_{strain}=\sum_{i,j,\sigma=\pm\frac{3}{2},\pm\frac{1}{2}}
\epsilon_{|\sigma|}^{s}a_{i,j,\sigma}^{\dagger}a_{i,j,\sigma}
\end{equation}
is the strain  Hamiltonian where $\epsilon_{|\sigma|}^{s}$ represents 
the strain-induced
energy with $\epsilon_{\frac{3}{2}}\not=\epsilon_{\frac{1}{2}}$.\cite{strain}
By adding strain, one may adjust the separation
between the HH and the LH bands.

\begin{figure}[htb]
\vskip-0.3cm
 \psfig{figure=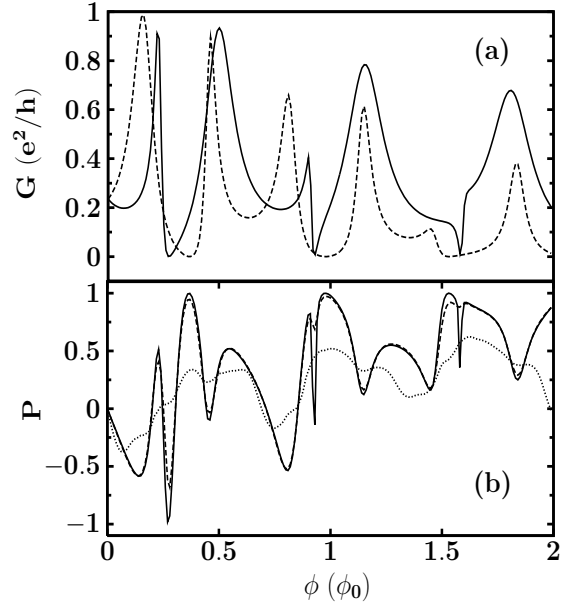,width=8.5cm,height=8cm,angle=0}
\vskip-0.3cm
  \caption{(a) Conductances of spin-up and -down HH's {\em vs.} the AB flux of 2D AB
frame when there is no LH channel in
 the leads.
Solid curve: $G^{\frac{3}{2}}$; Dashed curve: $G^{-\frac{3}{2}}$.
(b) Spin polarizations of HH averaged over 100 random disorder 
configurations {\em vs.} the 
AB flux with the disorder strengths
$W=0.1|t|$ (Dashed curve) and $W=0.5|t|$ (Dotted curve). 
The disorder-free case  is also plotted as solid curve.}
\label{fig:filter}
\end{figure}
The Spin-dependent conductance is calculated using the Laudauer-B\"{u}ttiker
formula\cite{Bu} with
the help of the Green function method.\cite{Da}
 The two-terminal spin-resolved conductance is given by
$G^{\sigma\sigma^{\prime}}=(e^2/h)\mbox{Tr}
[\Gamma_{1}^{\sigma}G_{1L}^{\sigma\sigma^{\prime}+}\Gamma_{L}^{\sigma^{\prime}}
G_{L1}^{\sigma^{\prime}\sigma-}]$ with $\Gamma_{1}(\Gamma_{L})$ representing the self-energy function for the
isolated ideal leads.\cite{Da} We choose the  perfect ideal Ohmic contact between the
leads and the semiconductor.  $G^{\sigma\sigma^\prime +}_{1L}$ and
$G^{\sigma\sigma^\prime -}_{L1}$ are the retarded and advanced Green functions
for the conductor, but with the effect from the leads included.

\begin{figure}[htb]
\vskip-0.3cm
\centerline{\psfig{figure=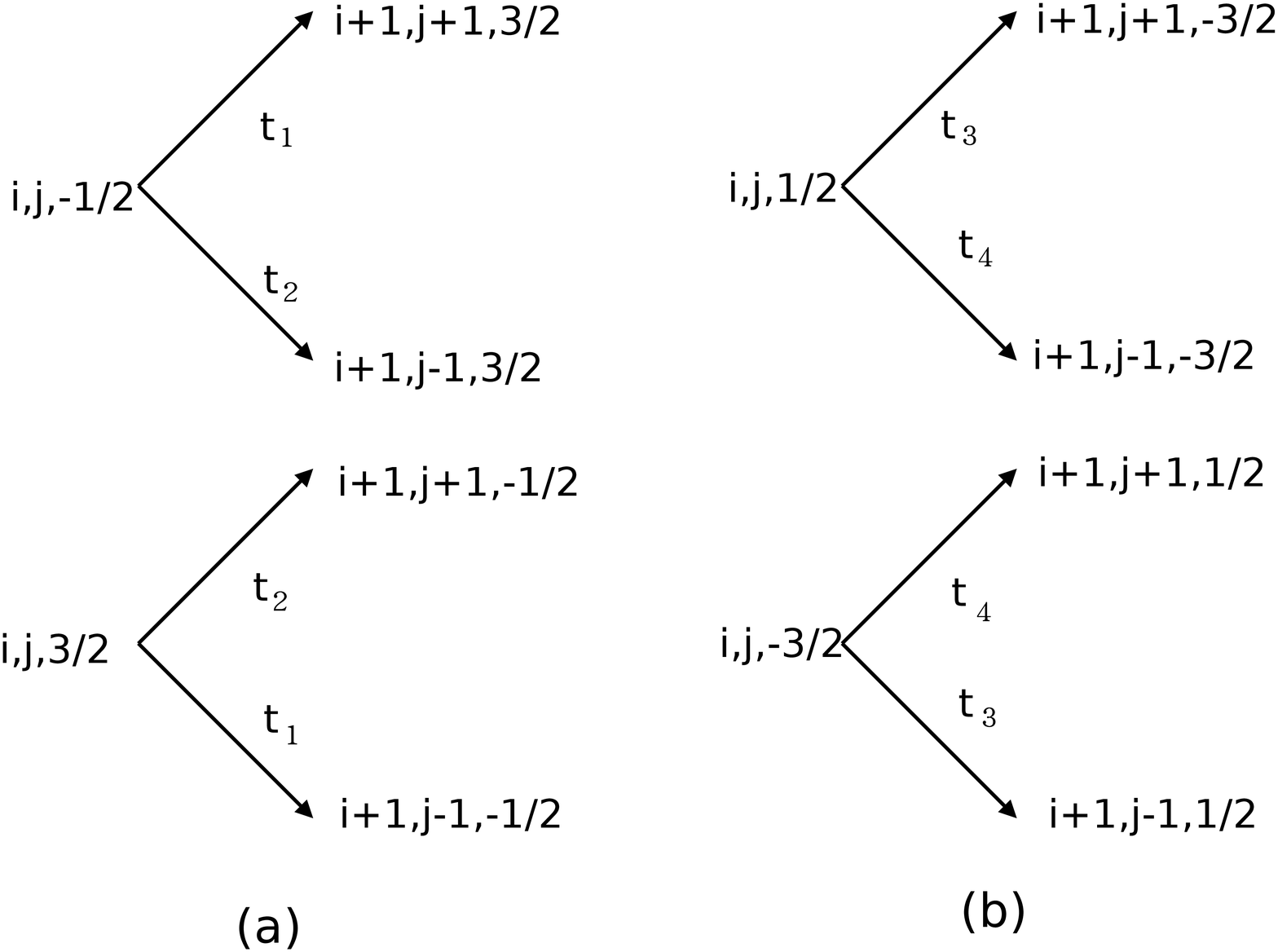,width=7cm,height=6cm,angle=0}}
\vskip -0.3cm
  \caption{(a) The next-nearest-neighbor  hopping of spin-down LH
$-\frac{1}{2}$ (spin-up HH $\frac{3}{2}$) from site $\langle i,j\rangle$ to $\langle
i+1,j\pm 1\rangle$ of subsystem $\alpha$;
(b) The next-nearest-neighbor  hopping of spin-up LH
$\frac{1}{2}$ (spin-down HH $-\frac{3}{2}$) from site $\langle i,j\rangle$ to $\langle
i+1,j\pm 1\rangle$ of subsystem $\beta$.}
\label{fig:shiyi}
\end{figure}

We perform a numerical calculation with $d=10a_0$, $W=40a_0$ and
$L=200a_0$. A hard wall potential is applied in the transverse
direction. In Fig.\ \ref{fig:filter}(a), we plot the conductances of the
spin-up HH
 $G^{\frac{3}{2}}=G^{\frac{3}{2}\frac{3}{2}}+G^{\frac{3}{2}-\frac{1}{2}}$
and the spin-down HH $G^{-\frac{3}{2}}=G^{-\frac{3}{2}\frac{1}{2}}+G^{-\frac{3}{2}
-\frac{3}{2}}$ at the right lead against the AB flux $\phi$.
It is noted that as there is no spin flip between the spin-up and -down HH's,
$G^{\frac{3}{2}-\frac{3}{2}}=G^{-\frac{3}{2}\frac{3}{2}}\equiv 0$.
By choosing a suitable strain on the two leads, one is able to
separate the HH and LH bands well apart
and consequently $G^{\frac{3}{2}-\frac{1}{2}}=
G^{-\frac{3}{2}\frac{1}{2}}=0$.
 We further align the HH band edges of the leads and the frame and
 choose a low Fermi energy which is $1.4|t|$ above the 
HH band edge  $E_{HH}^{0}$. As there is no strain applied on the
frame, the band edge of the LH is $8.29|t|$ above the 
$E_{HH}^{0}$. Therefore, the LH can not provide a real transport channel 
but only provides a virtual one.
One can see from the figure that when $B=0$, the
 conductances of the spin-up and -down HH's are identical. However
when $B\not=0$, the conductances vary differently with the AB flux.
The reason is understood as follows: When $B=0$, the phase of the subsystem
$\alpha$ which comes from the Luttinger spin-orbit coupling has
the same magnitude and the same sign as that of the
subsystem $\beta$ when a hole travels through
different arms as $H_{\alpha}(k_x,k_y)=H_{\beta}(k_x,-k_y)$. 
For instance, if one considers a
hole of subsystem $\alpha$ travelling through the upper arm along 
an arbitrarily chosen path $P_1$
which consists of a series of hopping, the phase that
 comes from the accumulation of the
imaginary part of the next-nearest-neighbor hopping, such as the hopping
in Fig.\ \ref{fig:shiyi}(a), is
the same as the phase that comes from a hole of subsystem
 $\beta$ travelling through
the mirror-symmetric path of $P_1$, {\em i.e.},
$P_1^\ast$. This is because that the hopping of subsystem
$\beta$ along $P_1^\ast$ consists of terms as shown in
Fig.\ \ref{fig:shiyi}(b) and one has $t_1=t_4$ and $t_2=t_3$ with
$t_1=-t_2=\frac{\sqrt{3}}{2}i\gamma_3t$.
Then it is obvious that the conductances of subsystems $\alpha$ and $\beta$
are exactly the same because the interferences of the two subsystems
that are determined by the summation of all the paths are 
entirely mirror-symmetric.
However, when $B\ne 0$, the phase shift from the AB effect destroys
the above symmetry, {\em i,e.} $t_1=\frac{\sqrt{3}}{2}i\gamma_3 V_{i+1,j+1;i,j}
\ne t_4=\frac{\sqrt{3}}{2}i\gamma_3V_{i+1,j-1;i,j}$ and
$t_2=-\frac{\sqrt{3}}{2}i\gamma_3 V_{i+1,j-1;i,j}\ne t_3
=-\frac{\sqrt{3}}{2}i\gamma_3V_{i+1,j+1;i,j}$.
Then the conductances of subsystems $\alpha$ and $\beta$ can be different.

\begin{figure}[htb]
\psfig{figure=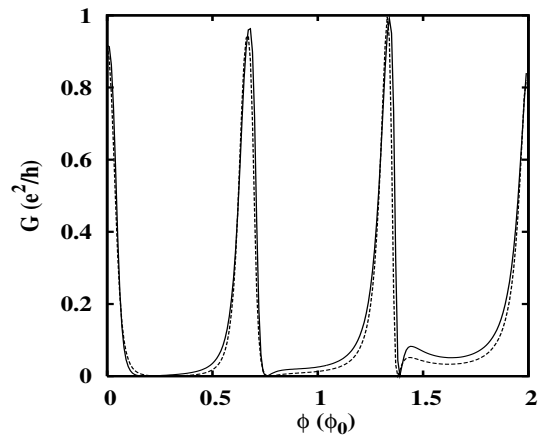,width=10cm,height=6cm,angle=0}
\vskip-0.3cm
\caption{Conductances of spin-up and -down HH's 
{\em v.s.} the AB flux of 2D AB frame when the LH band
edge is further increased by $50|t|$ from the case in 
Fig.\ \ref{fig:filter}.
Solid curve: $G^{\frac{3}{2}}$; Dashed curve: $G^{-\frac{3}{2}}$.}
\label{fig:far}
\end{figure}

It is interesting to see that although 
there is {\em no} real LH channel  in the frame
and the leads available for the transport
due to the low Fermi energy of the leads, the LH states still provide
virtual channels which  manifest different phases  
of the $\alpha$ and $\beta$ subsystems. It is due to the presence
of  these virtual channels that separate the HH's of different spins.
If one applies a strain to further increase the separation of the 
HH and LH in the frame of the above structure, then the contribution from the
virtual channels is suppressed and the spin separation becomes smaller.
This is demonstrated in Fig.\ \ref{fig:far} where
we use the same conditions as those in 
Fig.\ \ref{fig:filter}(a) except  the LH band edge
is lifted by $50|t|$. 
One can see that the difference of the conductance
 $G^{\frac{3}{2}}$ and $G^{-\frac{3}{2}}$ 
becomes much smaller. And when we lift the LH band edge even higher, 
 we find that $G^{\frac{3}{2}}=G^{-\frac{3}{2}}$ recovers the
ordinary AB effect in electron systems.

It is further seen from
Fig.\ \ref{fig:filter}(a) that by using the different conductances of the
spin-up and -down HH's at different AB flux, one is able to
make a spin filter. For example,
when $\phi\approx 0.4\phi_0$,
one can get spin-up polarization of the HH because
$G^{-\frac{3}{2}}=G^{-\frac{3}{2}-\frac{3}{2}}\approx 0$ and
 $G^{\frac{3}{2}}=G^{\frac{3}{2}\frac{3}{2}}\gg G^{-\frac{3}{2}}$;
 when $\phi\approx 0.3$, one can get  spin-down polarization of the HH
as $G^{\frac{3}{2}}\approx 0$ and $G^{-\frac{3}{2}}\gg G^{\frac{3}{2}}$.
In order to check the roubustness of this filter, 
we plot in Fig.\ \ref{fig:filter} (b) the spin polarization which is defined 
as $P=(G^{\frac{3}{2}}-G^{-\frac{3}{2}})/(G^{\frac{3}{2}}+G^{-\frac{3}{2}})$ 
averaged over 100 random configurations for different
disorder strengths versus the AB flux when Anderson disorder is considered.
 One can see that even when 
the disorder strength is $0.5|t|$, there is still large spin polarization.

\begin{figure}[htb]
\psfig{figure=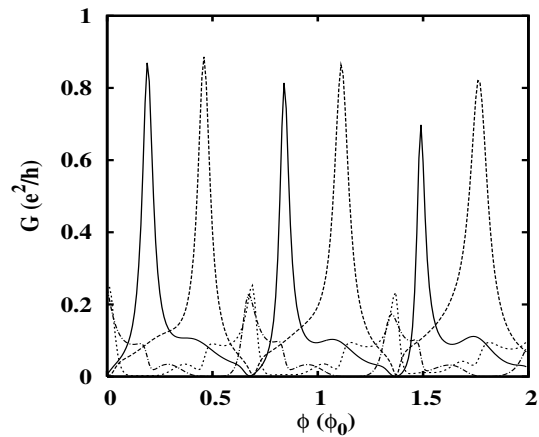,width=10cm,height=6cm,angle=0}
\vskip-0.3cm
\caption{Conductances of spin-up and -down HH's and LH's 
{\em v.s.} the AB flux of 2D AB frame.
Solid curve: $G^{\frac{3}{2}}$; Dashed curve: $G^{-\frac{3}{2}}$; Dotted curve: $G^{\frac{1}{2}}$; Chain
curve: $G^{\frac{1}{2}}$.}
\label{fig:HHinject}
\end{figure}

We further show that this structure can be used to generate spin polarized
current of LH while driving a spin-unpolarized 
HH charge current into the 2D frame.
This can be realized by applying a strain  on the left lead
which separate the HH and LH far away from each other
 and a different strain on the
frame to recover the $\Gamma$-point degeneracy. There is no strain on
the right lead. Therefore the LH band edge of the right lead is 
$8.29|t|$ above the HH band edge  $E_{HH}^{0}$ and the HH (and of course 
the LH) band edge of the frame is $13.6|t|$ above  $E_{HH}^{0}$. 
By applying a gate voltage on the left lead, 
one may align the  HH band edge of the left lead to be the same
as  $E_{HH}^{0}$.
The Fermi energy is chosen to be $14.65|t|$ above $E_{HH}^{0}$, {\em i.e.},
$1.05|t|$ above the HH (LH) band edge of the frame and
$6.36|t|$ above the LH band edge of the right lead.
Therefore, only a HH charge current can be injected from the left
lead into the frame but both HH and LH bands of the frame and the right
lead contribute to the transport.
The conductances of the HH and the LH are plotted versus 
the AB flux  in Fig.\ \ref{fig:HHinject}. It is seen from the
figure that  when $\phi=0.68\phi_0$ ($1.37\phi_0$), 
$G^{\frac{3}{2}}\approx 0$,
$G^{-\frac{3}{2}}\approx 0$ and $G^{\frac{1}{2}}=
G^{\frac{1}{2}-\frac{3}{2}}=0.294e^2/h$ ($0.371e^2/h$),
$G^{-\frac{1}{2}}=G^{-\frac{1}{2}\frac{3}{2}}=0.196e^2/h$ ($0.143e^2/h$).
Hence one can obtain a spin polarized current of 
LH with polarization $P=20$\ \% (44\ \%) where $P$ is defined as 
 $P=(G^{\frac{1}{2}}-G^{-\frac{1}{2}})/(G^{\frac{1}{2}}+G^{\frac{1}{2}})$.
Therefore, the AB effect of such a 2D frame provides us another 
scheme for spin filtering that a spin unpolarized HH can
be changed to the spin polarized LH.
Similarly by choosing $\phi=0.21\phi_0$
and $0.44\phi_0$,
one may get spin polarized HH current with a spin-unpolarized HH 
charge injection.
However, the energy dependence of this filter is very sensitive.
When we include the effect of disorder, 
the spin polarized LH current is always accompanied by the HH current. 

In conclusion, the AB effect in two-dimensional mesoscopic 
hole system is studied. 
We propose some schemes for spin filter 
based on the abundant interference characteristics.
When the band edges of the HH and LH are separated due to the 
confinement and the Fermi energy is lower than the LH band edge
but above the HH band edge,
we show that the LH still provides a virtual channel which 
leads to different phases for the spin-up and -down HH and 
gives rise to the spin separation. 
Therefore one can use the frame as a spin filter 
of HH by controlling the AB flux. 
Another spin filter is proposed when
a suitable strain is applied on the frame in order to make the band edges of
the HH and the LH close to each other and the channels of the leads are 
tuned so that both the HH and LH of the right lead but only the
HH of the left lead are below the Fermi energy. 
When a spin unpolarized HH current from the left lead is injected
into the frame, a spin polarized LH (or HH) current can be obtained by
controlling the AB flux. It is shown that the first scheme for
spin filter is very
robust against disorder whereas the second one is very poor against 
disorder.

This work was supported by the Natural Science Foundation of China
under Grant Nos. 90303012 and 10247002, the Natural Science Foundation
of Anhui Province under Grant No. 050460203 and SRFDP.


\begin{thebibliography}{10}

\bibitem{prinz} G. A. Prinz, Phys. Today {\bf 48}, 58 (1995);
 {\em Semiconductor Spintronics and Quantum
  Computation}, eds. D. D. Awschalom, D. Loss, and N. Samarth
  (Springer, Berlin, 2002); I. \v Zuti\'c, J. Fabian, and S. Das Sarma,
Rev. Mod. Phys. {\bf 76}, 323 (2004).
\bibitem{filter}M. J. Gilbert and J. P. Bird, Appl. Phys. Lett. {\bf
  77}, 1050 (2000); G. Papp and F. M. Peeters, Appl. Phys. Lett. {\bf
 78}, 2148 (2001); J. C. Egues ,C. Gould, G. Richter, and L. W. Molenkamp, Phys.
 Rev. B {\bf 64},
195319 (2001); Takaaki Koga, Junsaku Nitta, Supriyo Datta, and Hideaki Takayanagi,
Phys. Rev. Lett. {\bf 88},
126601 (2002); J. Fransson, E. Holmstr\"om, I. Sandalov, and O. Eriksson, Phys.
Rev. B {\bf 67}, 205310 (2003);
X. F. Wang and P. Vasilopoulos, Appl. Phys. Lett. {\bf 80}, 1400
  (2002); {\bf 81}, 1636 (2002).
\bibitem{stub} F. Sols, M. Macucci, U. Ravaioli, and Karl Hess,
  Appl. Phys. Lett. {\bf 54}, 350 (1989).
\bibitem{wu} J. Zhou, Q. W. Shi, and M. W. Wu, Appl. Phys. Lett. {\bf
  84}, 365 (2004).
\bibitem{wu1}M. W. Wu, J. Zhou, and Q. W. Shi, Appl. Phys. Lett. {\bf
  85}, 1012 (2004).
\bibitem{wu2}Q. W. Shi, J. Zhou, and M. W. Wu, Appl. Phys. Lett. {\bf
  85}, 2547 (2004).
\bibitem{ab}Y. Aharonov and D. Bohm, Phys. Rev. {\bf 115}, 485 (1969);
D. Frustaglia, M. Hentschel, and K. Richter, Phys. Rev. Lett. {\bf 87}, 256602
(2001); M. Popp, D. Frustaglia, and K. Richter, Nanotechnology {\bf 14}, 347 (2003).

\bibitem{trebin}H. R. Trebin, U. R\"{o}ssler, and R. Ranvaud,
Phys. Rev. B {\bf 20}, 686 (1979).
\bibitem{strain}G. L. Bir and G. E. Pikus, {\em Symmetry and Strain-Induced Effects
in Semiconductors} (Wiley, New York, 1974).
\bibitem{Bu} M. B{{\"u}}ttiker, Phys. Rev. Lett. {\bf 57}, 1761 (1986).
\bibitem{Da} S. Datta, {\em Electronic Transport in Mesoscopic
    Systems} (Cambridge University Press, New York, 1995).
\end{thebibliography}
\end{document}